\documentclass[longbibliography]{appolb}
\usepackage{graphicx}
\usepackage{hyperref}
\usepackage{amsmath}

\def\beq{\begin{equation}}
\def\eeq{\end{equation}}
\def\bea{\begin{eqnarray}}
\def\eea{\end{eqnarray}}
\def\beqn{\begin{eqnarray}} 
\def\eeqn{\end{eqnarray}}

\def\ln#1{\mathrm{log}\left(#1\right)}
\def\lb{\boldsymbol{\ell}}
\def\qon#1{q_{#1,0}^{(+)}}
\def\ii{\imath 0}
\def\uv{{\rm UV}}

\begin{document}
\title{QUANTUM ALGORITHMS IN PARTICLE PHYSICS
\thanks{Presented at the XLV International Conference of Theoretical Physics "Matter to the Deepest", Ustro\'n, Poland, 17-22 September, 2023.}%
}
\author{Germán Rodrigo
\address{Instituto de F\'{\i}sica Corpuscular \\ Universitat de Val\`{e}ncia -- Consejo Superior de Investigaciones Cient\'{\i}ficas \\
Parc Cient\'{\i}fic, E-46980 Paterna, Valencia, Spain.}}

\maketitle

\begin{abstract}
We motivate the use of quantum algorithms in particle physics and provide a brief overview of the most recent applications at high-energy colliders. In particular, we discuss in detail how a quantum approach reduces the complexity of jet clustering algorithms, such as anti-$k_T$, and show how quantum algorithms efficiently identify causal configurations of multiloop Feynman diagrams. We also present a quantum integration algorithm, called QFIAE, which is successfully applied to the evaluation of one-loop Feynman integrals in a quantum simulator or in a real quantum device. 
\end{abstract}

\section{Introduction}
\label{sec:intro}

Quantum algorithms have emerged recently as a promising avenue to efficiently tackle complex problems in the field of particle physics~\cite{Delgado:2022tpc}. Events occurring at high-energy colliders, such as the CERN's Large Hadron Collider (LHC), are typically analysed by factoring them according to the characteristic energy scale of their subprocesses. In this respect, recent applications of quantum algorithms in particle physics have considered specific aspects of the collision. For example, quantum algorithms have been explored for track reconstruction~\cite{PhysRevD.105.076012,Duckett:2022ccc,Schwagerl:2023elf} and jet clustering~\cite{PhysRevD.101.094015,Pires:2020urc,deLejarza:2022bwc,deLejarza:2022vhe}, including analysing the formation of jets in a medium ~\cite{Barata:2021yri,Barata:2022wim,Barata:2023clv}. Applications include also the simulation of parton showers~\cite{PhysRevLett.126.062001,PhysRevLett.127.212001,PhysRevD.106.056002}, quantum machine learning~\cite{Guan_2021,Wu_2021,cite-key} and the determination of parton densities~\cite{PhysRevD.103.034027}. On the most theoretical side, quantum algorithms have been used to evaluate helicity amplitudes~\cite{Bepari:2020xqi} and the colour algebra ~\cite{Chawdhry:2023jks} of elementary processes, and have been applied in the quest for selecting the causal configurations of multiloop Feynman diagrams~\cite{Ramirez-Uribe:2021ubp,Ramirez-Uribe:2022gxz,Clemente:2022nll,Sborlini:2023mws}. This concerns also quantum integrators~\cite{Herbert:2021xgs,AGLIARDI2022137228,deLejarza:2023qxk}, including their application to loop Feynman integrals~\cite{deLejarza:2024pgk}.

As quantum computers continue to advance, further applications are expected to appear offering a more complete picture. In general, it is assumed that the main advantage of a quantum approach is the potential speedup gain and the possibility of solving problems whose complexity scales exponentially or superpolynomially. These problems would become intractable at some point with classical computers. While this is true if the quantum principles of superposition and entanglement are exploited in the design of the quantum algorithm, it is also true that the current development of quantum technologies and hardware devices is still very limited due to the low coherence time of qubits and the inevitable quantum noise.  

The most interesting aspect of quantum algorithms in particle physics, in my opinion, is related to the original motivation of Richard P. Feynman~\cite{Feynman:1981tf}, suggesting that quantum effects, in this case particle collisions, should be better simulated with a quantum system. 

This presentation is devoted to describe in detail three recent applications of quantum algorithms in high-energy physics. Jet clustering in Section~\ref{sec:jet}, the bootstrapping of the causal representation of multiloop Feynman diagrams in the loop-tree duality in Section~\ref{sec:causal}, and the quantum integration of loop Feynman integrals, in Section~\ref{sec:loopintegration}.

\section{Jet clustering}
\label{sec:jet}

Clustering is one of the most frequent classical problems in many fields, such as machine learning and computational geometry.  In particular, jet reconstruction in particle physics is fundamental in the majority of experimental analyses. The most widely used jet clustering algorithm at the LHC is anti-$k_T$~\cite{Cacciari:2008gp}, which corresponds to the class of hierarchical or sequential jet recombination algorithms. In this type of clustering algorithm a distance is defined and at each step of the clustering process the pair of particles separated by the smaller distance is recombined into a new pseudoparticle until all pairs of pseudoparticles are separated by a distance greater than a minimum reference distance. 

Having to determine at each stage of clustering what is the absolute minimum distance between all pairs of particles is computationally expensive and the computation time increases rapidly with the multiplicity of the event. The computational 
complexity of the classical anti-$k_T$ is ${\cal O}(N^3)$, where
$N$ is the number of particles to cluster. However, the {\tt FastJet}~\cite{Cacciari:2011ma} implementation reduces this complexity to ${\cal O} (N\ln{N})$ by identifying each particle's geometrical nearest neighbour and optimizing the clustering by means of Voronoi diagrams. 

\begin{figure}[t]
\begin{center}
\includegraphics[scale=.30]{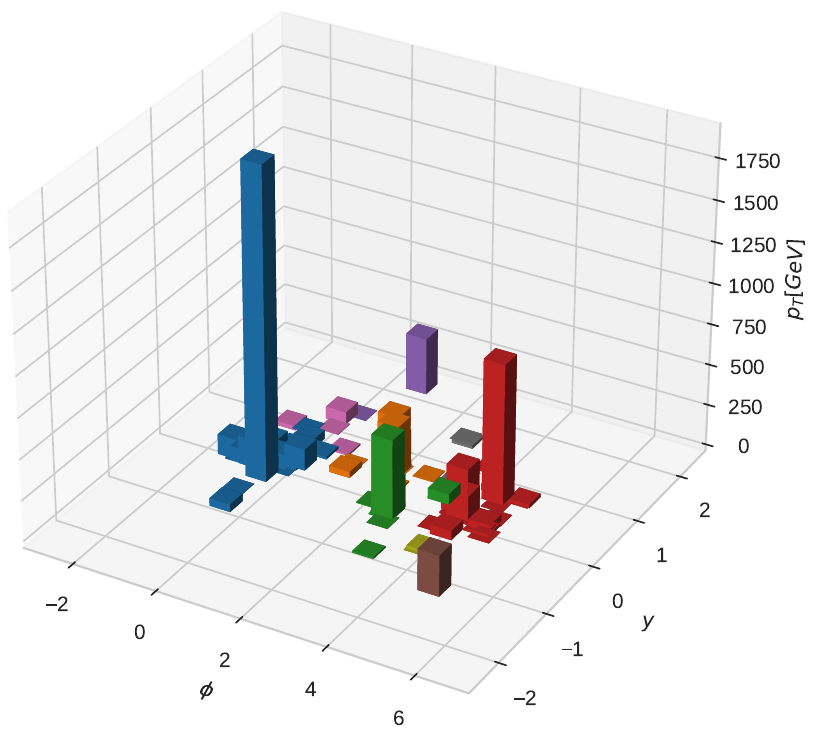}
\includegraphics[scale=.30]{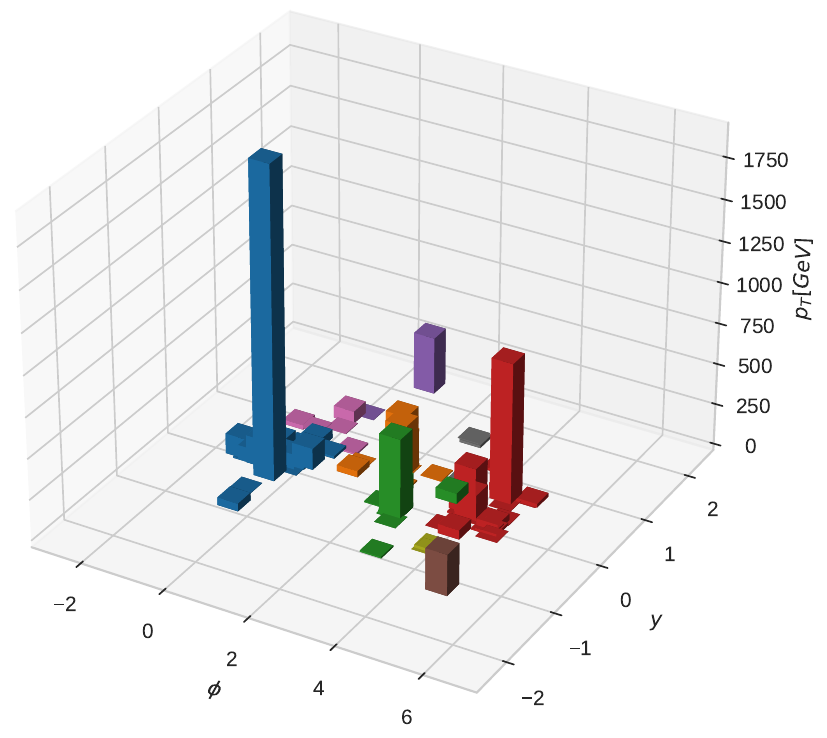}
\caption{Jet reconstruction of an event in the classical (left) 
and quantum (right) versions, respectively, of the anti-$k_T$
jet clustering algorithm with radius $R=1$.
\label{fig:jetclustering}}
\end{center}
\end{figure}

A quantum version of anti-$k_T$ with simulated LHC data has been 
presented in Refs.~\cite{deLejarza:2022bwc,deLejarza:2022vhe}. Without any optimization, this quantum counterpart of anti-$k_T$ would require ${\cal O} (N^2 \ln{N})$, and could achieve the same complexity as {\tt FastJet} by applying just geometric nearest-neighbour optimization. A comparison of the classical and quantum versions of anti-$k_T$ is illustrated in Fig.~\ref{fig:jetclustering}, where one can hardly appreciate any difference. There is, however, a fundamental difference between them as a quantum algorithm determines the minimun distance in a probabilistic way. Consequently, while the classical version of anti-$k_T$ is deterministic, i.e., given a set of particle data the clustering ordering will always be the same, in the quantum version the clustering ordering may be altered. This probabilistic determination of the mininum distance reduces the complexity of the algorithm. 
The final jet reconstruction is nevertheless quite similar because of the prevalence of collinear radiation as predicted by QCD. 
In Ref.~\cite{deLejarza:2022bwc}, we have also presented quantum versions of $k_T$~\cite{Catani:1993hr,Ellis:1993tq} and
of the general purpose clustering algorithms {\tt K-means}~\cite{macqueen1967some,ball1967clustering} and {\tt Affinity Propagation}~\cite{Frey2007ClusteringBP}.

\section{Causal configurations of multiloop Feynman diagrams}
\label{sec:causal}

A Feynman propagator describes the propagation of a particle between two interaction points in space-time in either directions. Therefore, we can formally write
\beq
G_{\rm F}(q_i) = \frac{1}{\sqrt{2}} \left( |0\rangle 
+ |1 \rangle \right)~,
\eeq
where the two propagation states are represented by 
$|0\rangle$ and $|1 \rangle$. The total number of states in a Feynman diagram is $2^n$, where $n$ is the total number of Feynman propagators. However, not all of these states are physical. Configurations in which a particle returns to the departure point, i.e., those configurations in which a particle describes a closed cycle, require traveling back in time and thus breaking causality.

In Feynman's representation, these nonphysical configurations are intrinsically present in the integrand and inevitably lead to nonphysical singularities. An unconventional approach is the loop-tree duality (LTD)~\cite{Catani:2008xa,Bierenbaum:2010cy} where a manifestly causal representation exists~\cite{Aguilera-Verdugo:2020set,Aguilera-Verdugo:2020kzc,Ramirez-Uribe:2020hes,JesusAguilera-Verdugo:2020fsn,Sborlini:2021owe,Sborlini:2023uyq,Ramirez-Uribe:2022sja,TorresBobadilla:2021ivx} and nonphysical singularities are absent, giving rise to numerically more stable integrads. On the one hand, the LTD representation fixes the directions of propagation allowed by causality. 
On the other hand, once the propagation directions are fixed, the manifestly causal LTD representation can easily be  bootstrapped. 

Given a multiloop Feynman diagram, our goal is to select from among all possible configurations obtained by identifying each Feynman propagator with a qubit, those configurations allowed by causality, and thus reconstruct its LTD representation. 
In this sense, there is an interesting connection with graph theory since the causal configurations correspond to directed acyclic graphs, which are commonly studied in many other fields.

\begin{figure}[t]
\begin{center}
\includegraphics[scale=.12]{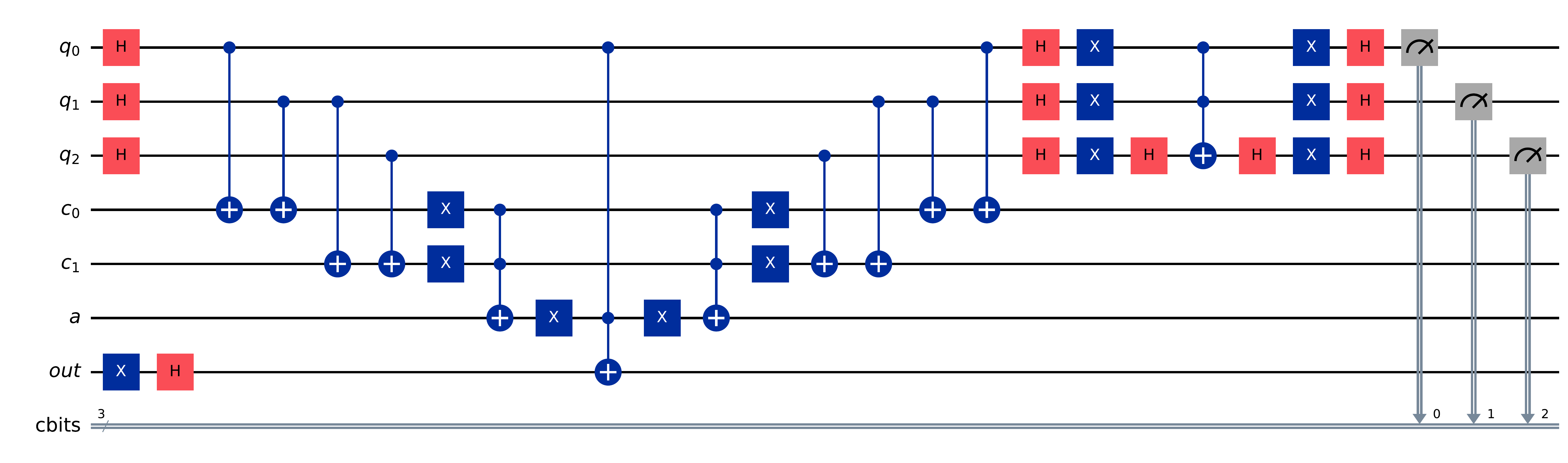}
\caption{Quantum circuit used to bootstrap the causal LTD representation of a three-vertex multiloop Feynman diagram
using amplitude amplification.
\label{fig:grover}}
\end{center}
\end{figure}

We have presented two different quantum algorithms to identify the causal configurations of a multiloop Feynman diagram. The first one is based on Amplitude Amplification or Grover´s algorithm~\cite{Ramirez-Uribe:2021ubp,Ramirez-Uribe:2022gxz}.
The second one, selects the causal configurations by minimizing a Hamiltonian in a Variational Quantum Eigensolver~(VQE) approach~\cite{Clemente:2022nll,Sborlini:2023mws}. Different energy levels of the Hamiltonian correspond to configurations with different number of cycles, and by construction, the
ground-state is directly associated to the subspace of
states related to directed acyclic graphs. 

In Fig.~\ref{fig:grover}, we present the quantum circuit used to select the causal configurations of a three-vertex multiloop Feynman diagrams in the Grover's approach. Note that the number of loops is not relevant because the only causal configuration of a group of propagators connecting two interaction vertices is the one in which all propagators are aligned in the same direction. The qubits $|q_i\rangle$, which are initially set in an uniform superposition, represent the three propagators. The ancillary qubits $|c_i\rangle$ and $|a_i\rangle$ are binary and loop clauses, respectively, that probe if a combination of qubits are in the same state. The qubit $|out\rangle$ is the Grover's marker. Measurements are done over classical bits. 

In both the VQE and Grover´s based algorithms, the main difficulty lies in the large number of causal configurations to be identified with respect to the total number of possible states, approximately half of all configurations, which requires several adaptations to implement the quantum algorithms efficiently. In both cases, we successfully identified all the causal configurations of Feynman diagrams with up to six interaction vertices, or four loops. In the case of the nonplanar diagram that appears for the first time at four loops, the $u$-channel~\cite{Ramirez-Uribe:2020hes}, it was necessary to resort on the quantum simulator Quantum Testbed~(QUTE) provided by Fundaci\'on Centro Tecnol\'ogico de
la Informaci\'on y la Comunicaci\'on~(CTIC) that supports simulations with~33 qubits. VQE requires less quantum resources than Grover’s algorithm (i.e. fewer qubits
and shorter circuits), still a larger number of executions
must be performed in order to achieve high success rates.

\section{Loop Feynman integration}
\label{sec:loopintegration}

The next steep after selecting the causal configurations of a loop Feynman diagram and constructing its LTD representation is to integrate over the loop momenta. To this purpose, we have 
proposed a new quantum integration algorithm called 
Quantum Fourier Iterative Amplitude Estimation~(QFIAE)~\cite{deLejarza:2023qxk} and applied this algorithm to benchmark one-loop Feynman integrals~\cite{deLejarza:2024pgk}, including its implementation in a real quantum device. 

The main novelty introduced by QFIAE consists of efficiently decomposing the integrand of the target function into its Fourier series using a Quantum Neural Network~(QNN). Then, each trigonometric component of the Fourier series is quantumly integrated using Iterative Quantum Amplitude Estimation~(IQAE)~\cite{Grinko:2021iad}. The Fourier decomposition encodes the target function with a minimum number of quantum arithmetic operations, and takes advantage of the fact that the trigonometric functions are more suitable for integration in a quantum approach. Moreover, the QNN is central to achieve a potential quadratic speedup. 

There are several reasons to choose the LTD representation. The absence of nonphysical singularities makes the integrand more stable numerically. The number of integration variables is independent of the number of external particles. The intregration domain is defined by the loop three-momenta, which is Euclidean. Finally, a local renormalisation of ultraviolet (UV) singularities easily renders the integrand UV finite. 

For example, the LTD representation of a locally renormalised one-loop bubble Feynman integral, with internal masses $m_1$ and $m_2$, and external momentum $p$, is~\cite{Aguilera-Verdugo:2020set}
\beq
{\cal A}^{(1,{\rm R})}_2(p,m_1,m_2) =  \int_{\lb} \left[\frac{1}{x_2} \left( \frac{1}{\lambda^+} 
+ \frac{1}{\lambda^-} \right)-\frac{1}{4(\qon{\uv})^{3}} \right]~,
\eeq
where  $x_2 = \prod_{i=1,2} 2 \qon{i}$, $\lambda^\pm = \sum_{i=1,2} \qon{i} \pm p_0$, and the on-shell energies are given by $\qon{i} = (\lb^2+m_i^2-\ii)^{1/2}$ with $i\in \{1,2\}$, assuming the external momentum has vanishing spatial components, $p=(p_0,{\bf 0})$. The last factor of the integrand that depends on $\qon{\uv} = (\lb^2+\mu_{\rm UV}^2-\ii)^{1/2}$ implements the local UV renormalisation, where $\mu_{\rm UV}$ is the renormalisation scale. The integration measure reads $\int_{\lb} \equiv \int d^3\lb /(2\pi)^3$.

Numerical results for several one-loop Feynman integrals using QFIAE in a quantum simulator have been presented in Ref.~\cite{deLejarza:2024pgk}, in full agreement with known analytic expressions. Moreover, for the one-loop tadpole integral, QFIAE was successfully implemented in real devices, including several strategies to mitigate quantum noise during execution. Deviations of less than $4\%$ were observed, which represents a noteworthy achievement for the current state of quantum computing technology.

\section{Conclusions}

Quantum algorithms represent a potential paradigm shift in computational methodologies for particle physics. Although still in its infancy, interest in quantum technologies in the field of high-energy physics is growing and several interesting applications covering different aspects of collider events have already emerged. We have discussed in detail three of these applications, which focus on jet reconstruction and loop Feynman integrals. At the moment, most of these applications can only be tested in quantum simulators, while their implementation in real quantum devices is limited by quantum noise. Besides a potential speedup, quantum algorithms in particle physics represent an appealing quantum way to analyse data and provide theoretical predictions. After all, quantum field theory is quantum by definition. 

\section*{Acknowledgements}

This work is supported by the Spanish Government (Agencia Estatal de Investigaci\'on MCIN/AEI/10.13039/501100011033) Grant No. PID2020-114473GB-I00, and Generalitat Valenciana Grants Nos. PROMETEO/2021/071 and ASFAE/2022/009 (Planes Complementarios de I+D+i, Next Generation EU). 

\bibliographystyle{JHEP}
\bibliography{2023_MTTD_rodrigo.bib}  

\end{document}